# On Curating Responsible and Representative Healthcare Video Recommendations for Patient Education and Health Literacy: An Augmented Intelligence Approach

Krishna Pothugunta
Accounting and Information Systems
Michigan State University
East Lansing, MI, USA
pothugun@msu.edu

Xiao Liu
Information Systems
Arizona State University
Tempe, AZ, USA
xiao.liu.10@asu.edu

Anjana Susarla
Accounting and Information Systems
Michigan State University
East Lansing, MI, USA
asusarla@msu.edu

Rema Padman
Information Systems and Public Policy
Carnegie Mellon University
Pittsburgh, PA, USA
rpadman@cmu.edu

## ABSTRACT

Studies suggest that one in three US adults use the Internet to diagnose or learn about a health concern. However, such access to health information online could exacerbate the disparities in health information availability and use. Health information seeking behavior (HISB) refers to the ways in which individuals seek information about their health, risks, illnesses, and health-protective behaviors. For patients engaging in searches for health information on digital media platforms, health literacy divides can be exacerbated both by their own lack of knowledge and by algorithmic recommendations, with results that disproportionately impact disadvantaged populations, minorities, and low health literacy users. This study reports on an exploratory investigation of the above challenges by examining whether responsible and representative recommendations can be generated using advanced analytic methods applied to a large corpus of videos and their metadata on a chronic condition (diabetes) from the YouTube social media platform. The paper focusses on biases associated with demographic characters of actors using videos on diabetes that were retrieved and curated for multiple criteria such as encoded medical content and their understandability to address patient education and population health literacy needs. This approach offers an immense opportunity for innovation in human-in-the-loop, augmented-intelligence, bias-aware and responsible algorithmic recommendations by combining the perspectives of health professionals and patients into a scalable and generalizable machine learning framework for patient empowerment and improved health outcomes.

## 1. Introduction

The World Health Organization (WHO) defines health literacy as "the cognitive and social skills which determine the motivation and ability of individuals to gain access to understand and use information in ways which promote and maintain good health"[23]. With estimates that only 12% of the US adult population is proficient in health literacy, healthcare knowledge and advice in video format may be more acceptable to much of the public[22].

Social media, particularly multi-media-rich visual social media, offers tremendous promise as a pathway for contextualized patient education and empowerment and public health literacy. However, how to produce fair and responsible recommendations in the form of curated health information videos that address the huge diversity of needs, abilities, and interests of consumers to improve their health knowledge, self-care skills and health outcomes has yet to be investigated. The challenge[23] with finding relevant and responsible advice on social media is two-fold: first, we need scalable and generalizable methods that can identify and retrieve health education videos with encoded medical content that is highly understandable. Second, we need a fair and bias-minimizing approach that ensures recommendations are not skewed against a particular demographic group or set of ideas. In this paper, we summarize both the challenges in this undertaking and the augmented intelligence approach we have developed that can subsequently be evaluated using social science methods.

### 1.1 Patient Education, Health Literacy, and Information Gaps

The current health education process for patients and the public has substantial information gaps. For patients engaging in searches for health information on digital media platforms, the gap in health literacy is exacerbated by their own lack of digital literacy, challenges of misinformation and disinformation, and the disparities in health information availability and use. While traditional media (such as books and brochures) or healthcare professionals have long been the primary source of health information, the advent of the Internet and social media has witnessed a sea change in health information seeking behavior (HISB). There is a dearth of knowledge on digital literacy gaps and content biases and how these are reflected in HISB on social media and digital platforms.

The evaluation of patients' comprehension of educational materials in the healthcare delivery setting is a challenge amplified by low health literacy levels to correctly interpret health information[21]. For patients to benefit from such educational materials also requires an elevated level of participation and engagement[10]. The rise of YouTube as a platform for the dissemination of healthcare information potentially offers a novel pathway to enhance patient education and the utilization of existing resources[16]. Users typically encounter videos on healthcare conditions through keyword searches on YouTube. It is a daunting challenge for both patients and clinicians to search for responsibly curated videos, retrieve them for each care delivery context, and use them in the form of just-in-time, contextualized, prescriptive digital therapeutic interventions.

Since users are heterogenous in their health information needs as well as in their levels of health literacy, relationship-oriented factors, such as trust and physician communication style, have been linked to disparities in patient satisfaction, delivery of preventive care services, appropriate use of referrals, and patient follow-through on treatments. Given this diverse set of challenges, it is necessary to develop a scalable, human-in-the-loop, augmented intelligence approach that synthesizes multiple machine learning methods with annotation by domain experts to extract relevant video content from digital platforms. Furthermore, a causal framework will permit us to understand the process of HISB and its impact on outcomes and evaluate the videos for accuracy and credibility to recommend for public use.

## 2. Methods

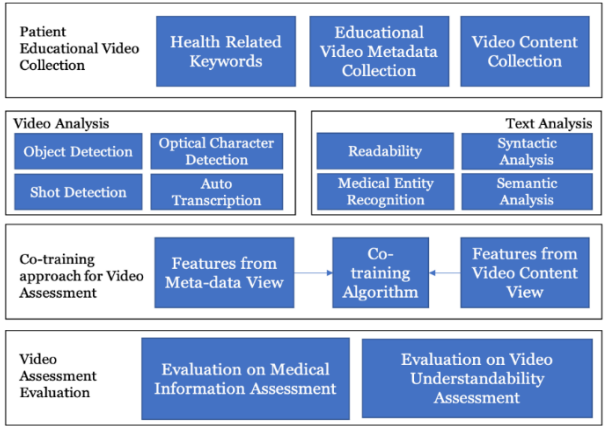

Figure 1. Medical Information and Video Understandability Assessment

Our prior work[11,12] focused on designing automated semi-supervised co-training methods to assess the medical information encoded in patient educational videos and the understandability of these videos from a patient education and health literacy perspective. Building on this prior research, we (1) assemble a corpus of diabetes-related YouTube videos, (2) develop an augmented intelligence approach to classify them using deep learning and natural language processing (NLP) methods according to whether they (i) contain relevant and medically valid information, and (ii) are understandable from a patient education perspective. In this preliminary study, we then examine age and gender bias of actors/narrators in the labeled video corpus that has been classified into four categories using the two criteria of high/low medical content and high/low understandability. Figure 1 summarizes the key algorithmic components of our prior work that results in the classified videos.

### 2.1 Video data collection on Diabetes

The focus on diabetes-related YouTube videos was motivated by its wide prevalence as a chronic condition and a contributing factor to many other serious health conditions, such as heart disease, stroke, nerve and kidney diseases, and vision loss. There is a critical need to educate patients and the public about the health risks and preventive measures for self-care and disease management. Understandable, multi-media and content rich educational videos can complement and support clinician and public health efforts.

Both individuals and diverse healthcare organizations have leveraged YouTube as a platform to offer thousands of videos on health education and instructions for understanding the symptoms and dealing with treatments[5]. The video collection process should represent what patients are searching for and what YouTube searches return to patients. Mimicking this process, we collect a large corpus of videos and their metadata on type 2 diabetes mellitus using YouTube Data API with the most significant 200 search keywords. We collect the first 50 videos for each search term and store the videos, their rankings, and metadata in a database for further analysis. The attributes we collect include channel ID (account name), publish time of the video, video title, video description, video tags, video duration, video definition, video caption availability, video rating, view count, like count, dislike count, and comment count. After removing duplicates and non-English language videos, the final corpus included 9,873 videos on a variety of diabetes related topics.

### 2.2 Video labeling

The ground truth for medical information and video understandability was established with a manual process[11]. A subset of the video corpus is labeled by medical professionals (domain experts) and graduate students (general consumers) for medical terminology and understandability using both UMLS semantic types and PEMAT criteria. The UMLS developed by the National Library of Medicine (NLM) maps an exhaustive lexicon of medical terms as concepts. The Agency for Healthcare Research and Quality's (AHRQ) Patient Education Materials Assessment Tool (PEMAT) is a systematic method with multiple criteria that can be used to evaluate and compare the understandability and actionability of patient education materials in written, audio or video format. Cohen's *kappa* statistic is used to measure concordance in the labels and consensus building discussions are used to resolve differences. The Kappa score for medical term annotation is 0.9 and interrater reliability for video understandability is 0.87.

### 2.3 Assessing understandability and medical information encoded in videos

Prior studies have relied on the judgment of domain experts such as health professionals to evaluate the medical information online[1]. Content rated by an expert (such as health professionals) is the most common approach to assessing the videos focused on health education. Health and medical websites are increasingly encouraged to apply for quality certificate assessments as proof of evidence that they are reliable sources of information. Medical information provided through a video can have several critical dimensions, such as content understandability by end users[20], volume of medical content, complexity of medical information provided[19] and so on. However, as the volume of online videos grows exponentially, using expert evaluation to assess all healthcare videos on YouTube is not a sustainable long-term solution.

We employ deep learning and natural language processing methods to characterize encoded medical information and understandability of the videos. Medical information in online videos is conveyed through medical terminology, which constitutes healthcare related words such as diseases and treatments. We employ deep learning to extract medical terminology embedded in a video[9]. We use the semantic types defined in UMLS to identify medical terms[10]. These terms can be classified as follows: diseases, treatments, tests, procedures, medical devises, medical professionals, and are used to label subset of the videos. We extend and train a Bidirectional Long Short-Term Memory (BiLSTM) model applied to the label videos to extract medical terms from the user-generated video descriptions at the sentence level[11,12].

We also apply the PEMAT criteria to assess understandability of each video. The understandability score is calculated by adding the scores for each feature/criterion in the tool and divided by total points possible (excluding the N/As). Given the challenges of literacy and designing visual aids to help comprehension of health information, a significant proportion of patient education materials are posited to be lacking on the dimension of understandability[8,16].

We develop two classifiers from two sufficient and conditionally independent views (i.e., video metadata and video content) and assess video understandability. Video data analysis forms the building blocks for designing a machine learning (ML) approach to evaluate patient educational videos. A large body of research has recently been devoted to enhancing performance of object detection within image frames. A multi–scale inference procedure that can create high–resolution object detection at a meager cost using Deep Neural Networks (DNNs)[18]. Convolutional Neural Network (CNN) algorithm overcomes the disadvantages associated with traditional machine learning algorithms on training large–scale image datasets. The use of unsupervised learning techniques has drastically improved due to the increased difficulty of acquiring labelled data[2]. State-of-the-art-performance for video mining tasks is often achieved by using one of many large open-source datasets or pre-trained models. Considerable advancements are made in Multi–Object Detection and Tracking (MODT), where an optimal Kalman filtering technique is used to track the moving objects in video frames[4].

In the video metadata view classifier, we leverage the features generated from video metadata to classify medical information and video understandability. Each video's metadata contains video title, description, and tags, which are submitted by the content creator. These elements suggest the purpose of a given video. We then implement a multi-view learning approach, a co-training model, to combine the results of the two classifiers.

The following steps summarize the co-training process: (1) An initial labeled dataset L and an unlabeled dataset U (rest of the data) are given; (2) The video metadata view of L is used to initialize the classifier $F_1$ (e.g., logistic regression), and the video content view of L is used to initialize the classifier $F_2$ (e.g., random forest). (3) $F_1$ and $F_2$ are used to predict the unlabeled data U. (4) A subset of the newly labeled positive ($p_1$ and $p_2$, resp.) and negative ($n_1$ and $n_2$, resp.) video samples with the highest confidence levels reported by $F_1$ and $F_2$, resp. is selected for comparison. When a video falls within the high confidence positive examples ($p_1$ and $p_2$) or the high confidence negative examples ($n_1$ and $n_2$), it indicates that this video is classified consistently positive or negative with high confidence. The video will then be added to the labeled dataset L. When a video gets inconsistent but high confidence predictions by $F_1$ and $F_2$, resp. (i.e., belonging to $p_1$ and $n_2$ or $n_1$ and $p_2$), the labelers will review the videos and provide supervision. The video with its label will be added to L. (5) Re-Train $F_1$ and $F_2$ based on updated L. The process is repeated until the unlabeled dataset is depleted or the optimal $F_1$ and $F_2$ is obtained (i.e., prediction performance is unchanged with additional training samples). Low confidence videos are discarded by the procedure.

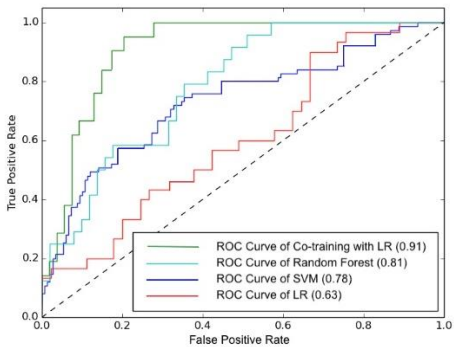

Figure 2. ROC Curve of Video Understandability Classification

The results of the co-training process provide the set of videos classified into high/low understandability and high/low medical information categories. The performance of video understandability classification is reported in Figure 2. The performance of medical information classification is reported in Table 1. These operations on the original video corpus are the initial key steps in building a responsible and reduced set of videos on diabetes that providers can feasibly evaluate further for accuracy and credibility. They also need further assessment for representativeness along multiple demographic and social determinants of health dimensions before recommending to patients and the public for education, engagement, and empowerment.

Table 1. Medical Information Classification Evaluation Results

|  | Precision | Recall | F1 |
|---|---|---|---|
| High Medical Information Videos | 88.7% | 83.9% | 86.2% |
| Low Medical Information Videos | 86.6% | 90.6% | 88.6% |
|  | **Overall Accuracy**: 87.5% | | |

## 2.4 Metrics for Representative Recommendations

Readability, content organization, and presentation are critical to healthcare consumers. These factors impact how patients engage with educational videos and whether the medical information can be delivered effectively. Our prior work focused on designing automated semi-supervised machine learning methods to assess the medical information encoded in patient educational videos and the understandability of these videos that consider patients' health literacy levels. Building on this, we conduct an ex-poste analysis of the labeled and classified videos to examine the presentation-level information such as the demographic characteristics of actors in the videos that may be insightful for assessing representativeness metrics through face detection[13]. Is there a change in the level of representativeness across the years in healthcare videos on YouTube platform?

Decision making on representativeness generally involves assessment at two levels: group and individual. Group representativeness checks whether a certain portion of population expressed by protected attributes (such as age, gender, and race) are treated fairly or not. Many experts have posited that the recommendation function in YouTube is a black-box, hence our study is going to concentrate solely on the post – processing aspect. More specifically, this preliminary analysis will check the fairness at the group level on both the age and gender attributes. The study can be further extended to the individual level fairness in the future.

We quantify bias faced by actors (i.e., people who are present in the videos who may be from different demographics. The gender of the presenter with no face visibility in the video is assessed through speech recognition. The age and gender of the presenters (i.e., presence of face) in the videos are identified using the following methods in Python: FaceLib and Pre-trained Caffe model. The source code for both the models are publicly available in GitHub repository[3,15,17]. The focus of the study will be on the videos with no / one actor. The study was intended to measure whether there is any impact of presenter's face visibility on the number of views generated by the video. If yes, what is the impact of the presenter's age / gender on the views generated?

H1: There is an impact of the presenter's face being visible on the views generated by the video

H2: There is an impact of the presenter's gender on the video views

The study also aims to investigate more about the relationship between the face visibility of the presenter and views contingent on the gender of the presenter.

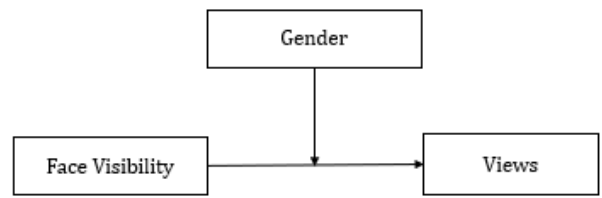

Figure 3: Model examining the moderator effect of presenter's gender on the relationship between presenter's face visibility and views

H3: The relationship between face visibility and views is moderated by gender, such that the relationship is strong for male presenters and weak for female presenters

Prior studies have shown major biases in machine learning applications employed such as facial recognition and so on. The accuracies of face – recognition systems used by US – based law enforcement is scientifically lower for personnel belonging to female group or between the ages of 18 and 30[7]. National Institute for Standards and Technology (NIST) also employed algorithms for gender detection that showed poor performance for female group[14]. There are three possible ways in which the unwanted bias can be mitigated in the machine learning pipeline: training data, learning procedure, and the predictions. These can eventually be classified into three classes: pre – processing, in – processing, and post – processing. We apply post-processing measures to the classified videos to generate some early insights into potential representativeness gaps in publicly available videos on diabetes on the YouTube platform.

## 3. Preliminary Results

A total of 1,675 videos related to diabetes is considered for the study. As shown in Table 2, the understandability and medical information scores for each of these videos are taken. Both the scores are based on binary values (0 = Low and 1 = High). A matrix of 2 * 2 (understandability * medical information) is created to check the variation of the number of actors in the videos, along with their age and gender.

Table 2. Summary Statistics of Videos

| Video Summary (Average) | Low MED | High MED | Low UND | High UND |
|---|---|---|---|---|
| #Videos | 552 | 1,123 | 320 | 1,355 |
| Views (Millions) | 22 | 38 | 14 | 29 |
| Subscribers (Thousands) | 142 | 155 | 125 | 82 |

*Note: MED – Medical Information, UND – Understandability*

As presented in Figure 4, the study did not include videos with more than one actor (280), as measuring the representativeness of multiple actors in a single video is complicated and left for further research. A few videos (133) that are not related to diabetes such as musical, religious, and so on are removed from the analysis. A small sample of videos (29) could not be read using the applied face recognition packages.

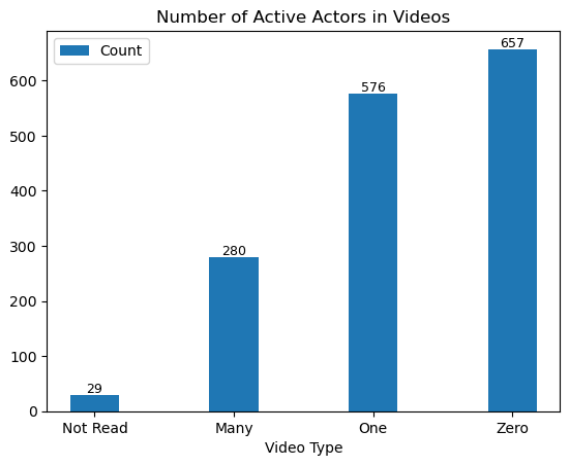

Figure 4. Video classification based on face recognition of actors.

Further analysis is conducted in the videos with zero / one actor to check the representation of gender attribute. As shown in Figure 5, videos with high understandability and medical information are significant portion of the sample, in which most of the videos in all four categories are dominated by male gender group. In the videos presented by female group for all subsections, the number of videos with no face presence is higher than the ones with presenter's presence, which is opposite in the case of male group of presenters. Few videos (84) that do not have any narration have been dropped.

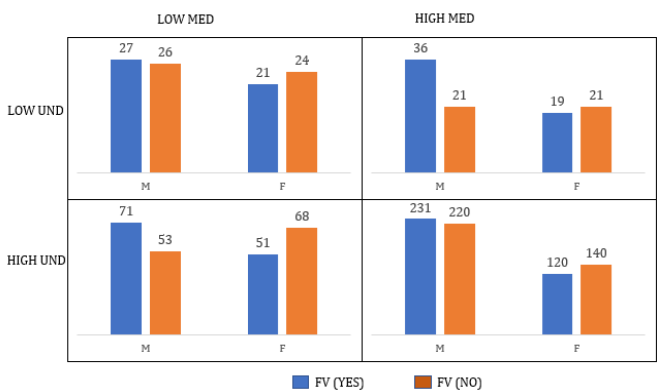

Figure 5. Descriptive statistics of understandability and medical information with gender of actor in videos with zero / one presenter; FV – Face Visibility of the presenter; Total videos = 1149.

The number of views generated by the videos seem to have a significant positive relationship with the videos with male presenters and videos with high understandability, as shown in Table 3.

Table 3. Correlation Matrix

| | 1 | 2 | 3 | 4 | 5 |
|---|---|---|---|---|---|
| 1.FV | 1 | — | — | — | — |
| 2.Gender | .079** | 1 | — | — | — |
| 3.MED | .004 | .101*** | 1 | — | — |
| 4.UND | -.025 | .031 | .204*** | 1 | — |
| 5.viewCount | -.022 | .113*** | .059* | .11*** | 1 |

*Note*: N = 1147; *p < 0.05, **p < 0.01, ***p < 0.001, two – tailed; Videos with zero views are removed (n = 2).

To verify the impact of face visibility and gender on the views generated GLM and LASSO regressions are conducted using GLM and GLMNET packages in R. The data is divided into train (70%) and test (30%) samples. Based on the results in Table 4, there was no significant impact on the presence on presenter's face in the videos, not supporting H1. However, the gender (GLM = .635, LASSO = .751) of the presenter has a significant impact on the views generated by the videos, supporting H2. Finally, the interaction effect shows that the relationship between the presenter's face visibility and the video views is not contingent on the presenter's gender, not supporting H3. Simple slope analysis also reveals similar results which shows that there is no interaction between face visibility and gender on the video views, as shown in Figure 6.

Table 4. Regression Results

| | GLM | LASSO | GLM | LASSO |
|---|---|---|---|---|
| 1.FV | -.179 | -.084 | -.173 | -.081 |
| 2.Gender | .635** | .751 | .594** | .741 |
| 3.FV: Gender | .014 | .000 | .031 | .000 |
| 4.MED | | | .161 | .000 |
| 5.UND | | | .731 *** | .506 |

*Note*: N = 1147; Best Lambda = 0.038; **p < 0.01, ***p < 0.001, two – tailed.

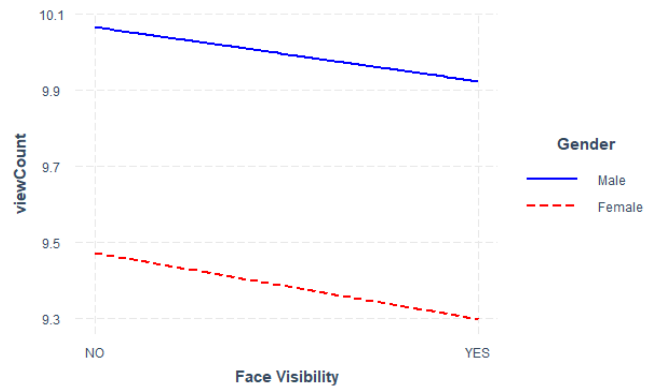

Figure 6. Simple slope analysis to highlight interaction effects.

The results highlight interesting findings for further research, which underlines the importance of understandability (GLM = .731, LASSO = .506) on the videos viewed on YouTube.

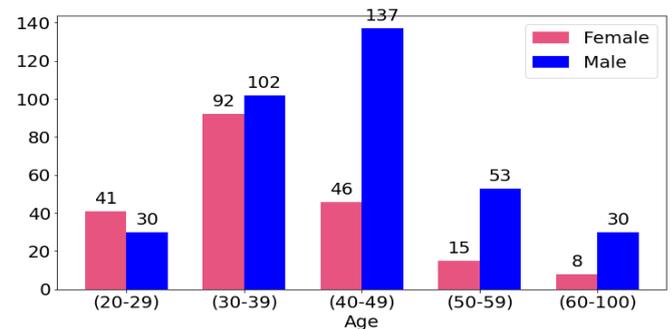

Figure 7. Descriptive statistics of age and gender break up of videos with one actor.

The research study can be extended to include the age also another protected attribute to measure the representativeness of the videos. As shown in Figure 7, actors within age groups between 30 and 50 years tend to present more videos, and the presence of female actors significantly decrease with age. The age detection can also be extended to videos with no actor presence, using speech recognition. These are preliminary observations from a limited set of videos on one condition (diabetes) and need to be examined further for larger video corpus and other conditions.

## 4. Discussion and Conclusions

Automating the easy retrieval of understandable patient education videos for clinicians to recommend to their patients has significant societal impact. Decision makers can eventually use the insights to develop best practices to promote health education videos for improving patient and public health literacy. We have identified three major criteria for success. First is the successful development of the augmented intelligence-based methods for selecting videos that can address existing health literacy divides. Second is the broader applicability of these methods for other areas such as establishing credible and authoritative sources of information for education, health messaging, etc. The third criterion is that of bias-sensitive recommendations. We intend to next examine if there are significant differences in terms of readability index whereby videos with higher readability that only account for a small proportion in data enjoy much higher recommendation quality than others.

Finally, we aim to build a video recommendation classifier using the dimensions of understandability and medical information highlighted earlier as well as new measures of algorithmic fairness by examining the discrepancy in what is recommended by the YouTube baseline and our method and compare them with the performance of our recommendations on various attributes. We can also conduct audits of fairness by looking at popular topics, compiling search queries and examining link stance from a bias/ fairness perspective as well as new m fairness and representativeness.

As the study employs pre – existing machine learning algorithms for face detection, there are a few limitations that need to be considered. The error rates for detecting women with darker skin is significantly higher when compared with women with lighter skin. Significant portion of the sample is generally drifted towards the population with lighter skin. However, cross – validation and several re – runs are conducted to reduce the errors in age and gender detection.

## ACKNOWLEDGMENTS

This study was partially supported by grant #R01LM013443 from the National Library of Medicine (NLM) of the National Institutes of Health (NIH). We also thank the graduate students in the Spring 2022 Capstone project at Carnegie Mellon University who assisted in the collection and the classification of videos.